# Bulk Superconductivity in $Fe_{1+y}Te_{0.6}Se_{0.4}$ Induced by Removal of Excess Fe

Wei Zhou[1], Yue Sun[1,2], Shuo Zhang[1], Jincheng Zhuang[1], Feifei Yuan[1], Xiong Li[1], Zhixiang Shi[1*], Tatsuhiro Yamada[2], Yuji Tsuchiya[2], and Tsuyoshi Tamegai[2]

*1. Department of Physics and Key Laboratory of MEMS of the Ministry of Education, Southeast University, Nanjing 211189, China*
*2. Department of Applied Physics, The University of Tokyo, 7-3-1 Hongo, Bunkyo-ku, Tokyo 113-8656, Japan*

**Abstract:** Experimental evidences from transport, magnetic, and magneto-optical (MO) image measurements confirmed that arsenic (As) vapor annealing was another effective way to induce bulk superconductivity with isotropic, large, and homogenous superconducting critical current density ($J_c$) in $Fe_{1+y}Te_{0.6}Se_{0.4}$ single crystal. Since As is an exotic and easily detectable heavy element to $Fe_{1+y}Te_{0.6}Se_{0.4}$ single crystal, As vapor annealing is very advantageous for the study of annealing mechanism. Detailed micro-structural and elemental analyses exclude the possibility that intercalating or doping effect may happen in the other post-annealing methods, proving that Fe reacts with As on the surface of the crystal and the reaction itself acts as a driving force to drag excess Fe out. The removal of excess Fe results in the good superconductivity performance.

1. Introduction

Since the observation of superconductivity around 8 K in $Fe_{1+\delta}Se$ ('11' system),[1] great efforts have been devoted to this compound for its simplest crystal structure among iron-based superconductors (IBSs)[2-4] and less toxic nature which is advantageous for applications. Partial substitution of Te for Se can effectively enhance the superconducting transition temperature ($T_c$) to around 14.5 K.[5-7] However, poor superconductivity performance (low zero-resistivity temperature, small superconducting shielding volume fraction, and low critical current density, etc.) was found for the as-grown (hereafter noted as AG) single crystal.[6, 8] Recently, many post-annealing methods were demonstrated to be effective in improving superconductivity performance, such as annealing in vacuum,[8, 9] in air,[10] in nitrogen,[11] and in $I_2$ vapor.[11, 12] Our results[13] verified that annealing the sample in $O_2$ atmosphere was effective to induce bulk superconductivity and pure $N_2$ annealing could not work. Furthermore, immersing the sample in acids or alcoholic beverages can only induce surface superconductivity. High-quality single crystals with large, homogeneous, and almost isotropic critical current density have also been obtained through accurate control of $O_2$ annealing condition.[14] It is well known that superconductivity in "11" system is extremely sensitive to the stoichiometry of Fe atoms[15] and the excess Fe located in the interstitial site in the Te/Se layer causes magnetic correlations and suppresses superconductivity.[16-18] Possibly, any method which can cause the reduction of the concentration of excess Fe is functional to improve the superconductivity performance. Actually, annealing the sample in the vapor of all chalcogen elements was proved to induce bulk superconductivity in the present material.[19, 20] Despite many of the post-processing methods mentioned above work well to obtain good superconducting properties, some fundamental issues remain to be addressed. To date, only a few detailed studies to clarify the compositional variation were reported. The underlying mechanism of the various effective post-annealing methods is still unclear.

In this paper, we report that annealing the AG $Fe_{1+y}Te_{0.6}Se_{0.4}$ single crystal in arsenic (As) vapor can also induce bulk superconductivity, which confirms the key role of excess Fe again. As vapor annealing has its advantage compared with the previous annealing processes, especially for the study of the annealing mechanism. Since As is an exotic and heavy element, one can easily detect its concentration and obtain the detailed information about the reaction with excess Fe. Our elemental analysis puts a strong evidence to manifest that As reacts with excess Fe on the surface of the crystal and the reaction itself may play a driving role in dragging excess Fe in the inner part out, reacting with As, and precipitating on the crystal's surface.

## 2. Experiments

Single crystals with a nominal composition $FeTe_{0.6}Se_{0.4}$ were grown using the similar method to the previous reports[8, 13]. The AG crystals were cleaved into thin slices and sealed together with proper amounts of As grains in silica tubes of a constant volume (inner diameter ~ 10 mm Φ, length ~ 135 mm). A turbo pump was used to vacuum the silica tubes to pressure lower than $8.5 \times 10^{-3}$ Pa simultaneously in the process of sealing. Such pressure obviously excludes the influence of $O_2$.[13, 21] The sealed silica tubes were put into a box-furnace and heated to 400 °C in 1h. After holding at 400 °C for 24h, the silica tubes were quenched into water at room temperature. Magnetic measurements were performed via the VSM option of a physical properties measurement system (PPMS). The typical sample size for the magnetization measurements is 2.4–3.1 mm × 1.8-2.6 mm × 20-50 μm. Resistivity measurements were carried out using the four-probe method on PPMS. Magneto-optical (MO) images were obtained by using the local field-dependent Faraday effect in the in-plane magnetized garnet indicator film employing a differential method.[22] Micro-structural and elemental analyses were done by field emission scanning electron microscopy (SEM) equipped with energy-dispersive X-ray spectroscopy (EDXS).

## 3. Results and Discussion

Figure 1 (a) shows the temperature-dependent zero-field-cooled (ZFC, closed symbols) and field-cooled (FC, open symbols) magnetization measured at 20 Oe for

$Fe_{1+y}Te_{0.6}Se_{0.4}$ single crystals annealed in As vapor with different molar ratio ($n$ = As : $Fe_{1+y}Te_{0.6}Se_{0.4}$). Diamagnetic signal was not observed for the AG and $n$ = 0.01 samples, while weak diamagnetism was induced when $n$ is increased up to 0.02. Significant improvement of superconductivity was obtained for As vapor annealed (hereafter noted as AVA) samples with molar ratio $n$ = 0.05 or higher. The evolution of superconducting transition temperature with molar ratio $n$ was shown in Fig. 1 (b). As can be seen, $T_c^{mag}$, determined by the separation point between ZFC and FC curves, is fast enhanced with increasing the molar ratio $n$ and keeps an almost constant value in the range of 0.05≤ $n$ ≤ 0.4. The As vapor annealing effect is noticeable. On the other hand, the annealing effect is also relevant to the annealing temperature. As shown in the inset of Fig. 1 (b), for the same molar ratio ($n$ = 0.1), sample annealed at 600 °C exhibits much worse superconductivity performance. The invalidity of 600 °C annealing was also noticed in other vapor annealing methods, which may be associated with phase decomposition (or certain phase transition) for long time annealing at high temperatures based on the comparison of X-ray diffraction patterns (not shown) between the AG and AVA (600 °C) samples. Fig. 1 (c) shows the magnetic hysteresis loops (MHLs) for the AG and AVA ($n$ = 0.2) samples at 5 K ($H \parallel c$). As can be seen, no superconducting loop was found in MHLs for the AG sample, while large hysteresis which presumably was induced by bulk superconductivity was observed for the AVA sample. Fig. 1 (d) provides the transport results of temperature-dependent resistivity ($\rho$-$T$) for the AG and AVA samples. For resistivity measurement, both surfaces (top and bottom) of the AVA sample ($n$ = 0.2) was peeled off. Only the inner part, shiny as the AG single crystal, was used to characterize the most intrinsic properties. A change from semiconducting to metallic $\rho$-$T$ behavior was witnessed at low temperature beyond $T_c$. The change from semiconducting to metallic normal state was also observed in other post-processing methods, which was explained by electron delocalization resulted from the removal of excess Fe.[14, 20, 23] Sharper superconducting transition was also found for the AVA sample, which consistently manifests the improvement of superconductivity.

In order to strengthen the idea that the good superconductivity performance induced by As vapor annealing is a bulk property, field-dependent magnetization measurements with the magnetic field applied both parallel to *ab*-plane and *c*-axis and magneto-optical image measurements were performed. As can be seen in Figs. 2, large hysteresis and typical fishtail effects (also known as "second peak" effect) were observed in MHLs for both two directions. Using the famous Bean model,[24] $J_c = 20\Delta M / a(1 - \frac{a}{3b})$, where $\Delta M$ (unit: emu/cm$^3$) is $M_{down} - M_{up}$, $M_{down}$ and $M_{up}$ are magnetization when sweeping fields down and up, respectively, calculated critical current density $J_c$ (unit: A/cm$^2$) was also plotted. Here *a* (unit: cm) and *b* (unit: cm) are sample width and length ($a < b$). The fishtail effect is more obvious for $H \parallel c$, which can be seen clearly in the $J_c(H)$ curves. With increasing the measured temperature, the second peak moves quickly to low field and becomes much more obvious for both directions. A large critical current density which easily exceeds $10^5$ A/cm$^2$ was obtained for the AVA sample in zero field. The obtained self-field critical current density, $J_c^{H \parallel ab}$ and $J_c^{H \parallel c}$ at 3 K are estimated to be $3.8 \times 10^5$ A/cm$^2$ and $3.3 \times 10^5$ A/cm$^2$, respectively. Both $J_c^{H \parallel ab}$ and $J_c^{H \parallel c}$ at 3 K are remarkably robust against magnetic field up to 9 T. Such a large and isotropic $J_c$ which was only obtained by few annealing methods[14, 20, 25] could not come from surface or filamentary superconductivity with a small superconducting component but must be the sign of bulk superconductivity.

Figure 3 shows the MO images of the As-annealed crystal in the remanent state at temperatures ranging from 5 to 13 K. This state is prepared by applying 400 Oe along the *c*-axis of crystal for 1s and removing it after zero-field cooling. Typical MO images at 5, 7, and 13 K are shown in Figs. 3 (a), (b), and (c), respectively. Clear roof-top patterns can be observed, which indicates the homogeneous current flow in the crystal. Such roof-top pattern is also observed in high-quality Fe$_{1+y}$Te$_{0.6}$Se$_{0.4}$ single crystals prepared by O$_2$ and Te annealing.[14, 20, 26] Fig. 3 (d) shows magnetic induction profiles along the dashed line in Fig. 3 (a) at different temperatures. Using an approximate formula for a thin sample, $J_c \sim B/d$, where $B$ is the trapped field in the crystal, and $d$ is the thickness of the crystal, $J_c$ at 5 K can be roughly estimated as $\sim 1.5 \times 10^5$ A/cm$^2$, similar to that obtained from MHLs. The large and homogeneous critical current density obtained by MHLs and MO strongly proves that As vapor annealing is effective to induce bulk superconductivity in Fe$_{1+y}$Te$_{0.6}$Se$_{0.4}$.

To investigate the underlying microstructure and compositional variation induced by As vapor annealing, detailed back-scattered electron image and energy-dispersive X-ray spectroscopy (EDXS) measurements were carried out for the AG and AVA samples. Some back-scattered electron images and corresponding detected elemental distribution are shown in Figs. 4 and table 1. All numbers in table 1 are averaged values of several measured points. It should be pointed out that, the brightness of the back-scattered electron image should somewhat reflect the distribution of elements. Figs. 4 (a) and (b), including the image extracted from EDXS measurement shown in the upper-right corner of Fig. 4 (a), manifest the relatively uniform elements distribution for both the AG sample and the AVA sample (inner part) whose double surfaces have been peeled off by an adhesive tape. Because the stability of the excess Fe is less on the surface and the Fe adopts more locations in the bulk of the sample, the detection data of Fe is slightly less.[27] Inductively-coupled plasma (ICP) spectroscopy measurements of similar AG crystals show an average composition of Fe$_{1+y}$Te$_{0.6}$Se$_{0.4}$ with $y \sim 0.14$.[28] Therefore, we rescaled the Fe concentration by multiplying $\eta$ which represents the ratio of the detected values between ICP and EDXS for the AG sample. In this manner, the average composition for the AG sample is Fe$_{1.14}$Te$_{0.61}$Se$_{0.39}$. For the inner part of the AVA sample, the element distribution was also homogeneous. However, the phase composition was changed into Fe$_{1.06}$Te$_{0.615}$Se$_{0.385}$ with reduced Fe concentration of $y \sim 0.08$. It is important to note that all 10-point measurements did not show any presence of As, which excludes the possibility of As doping or intercalation in the crystal after As vapor annealing. Similarly, we can guess that doping and intercalation do not occur during other post-annealing processes, such as chalcogen vapor annealing.[19, 20] Considering the remarkably different electronic behaviors reflected in $\rho$-$T$ curves for the AG and AVA sample, and no trace of As in the inner part of AVA sample, there is no doubt that the motion of excess Fe actually occurred during the annealing process.

Figure 4 (c) shows the back-scattered electron images of the AVA sample without intentionally cleaving down the surface. Rough and nonuniform surface, which originates from the random formation of different cleavage planes after annealing, was observed for the AVA sample. An enlarged view was shown in Fig. 4 (d). Three distinct areas (A, B, and C) can be witnessed. The composition of area A is $Fe_{1.03}Te_{0.59}Se_{0.41}$, very close to that of the inner part of the AVA sample. In area B, tree-like rimous surface was observed. The composition was changed into $Fe_{0.73}Te_{0.60}Se_{0.40}As_{0.07}$, indicating that area B is an over-annealing region with huge decrease of Fe concentration. The trace amount of As may be caused by the slightly remnant $FeAs_2$ that will be discussed in the following. Area C is the outermost surface and the elemental composition is $FeTe_{0.08}As_{1.86}$, indicating new formation of $FeAs_2$ layer (also confirmed by X-ray diffraction pattern ) on the surface of the crystal.

Next, let's pay more attention to the Fe concentration variation caused by As vapor annealing. As can be seen in table 1, from inner part (b) of the AVA sample to area A, and further to area B, continuous decrease in the Fe concentration was found. We believe that more Fe was removed from the outer region and this takes the full responsibility for the different *M-H* results between the inner part of annealed crystal and the whole annealed sample in ref. [13]. Naturally, one can suppose that the reaction between Fe and As formed a new $FeAs_2$ layer precipitating on the surface, which also resulted in a dissipation layer (regions like area B) and a concentration gradient of Fe. Because of the concentration gradient, excess Fe diffused from the inner part of crystal to the surface. The situation may be also similar to the other post annealing methods reported before.

## 4. Conclusions

We have studied effects of arsenic vapor annealing on the superconductivity in $Fe_{1+y}Te_{0.6}Se_{0.4}$ single crystal through transport and magnetic measurements including magneto-optical imaging. Arsenic vapor annealing was demonstrated to be an effective way to obtain bulk superconductivity with an isotropic, large, and homogenous $J_c$ in Fe$_{1+y}$Te$_{0.6}$Se$_{0.4}$ single crystal. Detailed micro-structural and elemental analyses manifest that excess Fe reacts with As on the surface and the reaction itself may act as a driving force to drag excess Fe out of the crystal. The removal of excess Fe from the inner part of the crystal via As vapor annealing results in bulk superconductivity in Fe$_{1+y}$Te$_{0.6}$Se$_{0.4}$ single crystal.

**Acknowledgments**: This work was supported by the Natural Science Foundation of China, the Ministry of Science and Technology of China (973 project: No. 2011CBA00105), and Jiangsu Science and Technology Support Project (Grant No. BE 2011027).

*E-mail: zxshi@seu.edu.cn

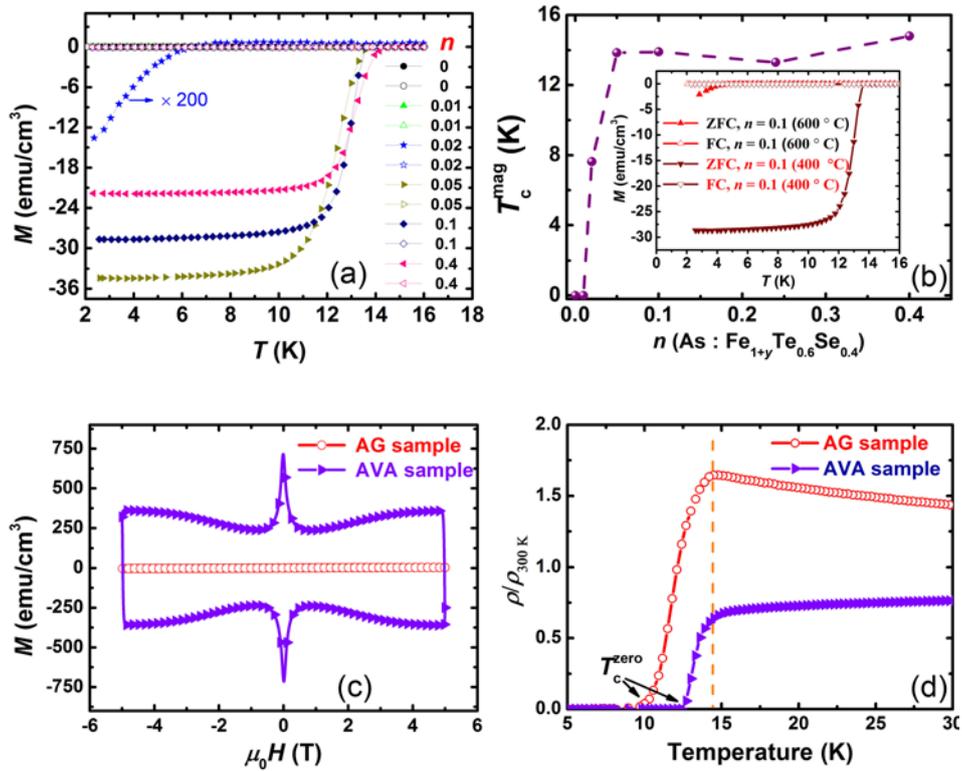

Fig. 1 (Color online) (a) Temperature dependences of zero-field-cooled (ZFC, closed symbols) and field-cooled (FC, open symbols) magnetization at 20 Oe ($H \parallel c$) for samples annealed with different amounts of As. $n$ ($n$ = molar (As)/molar ($Fe_{1+y}Te_{0.6}Se_{0.4}$)) is the molar ratio of As to $Fe_{1+y}Te_{0.6}Se_{0.4}$. For the AG sample, $n = 0$. For the AVA sample with $n = 0.02$, the magnetization for ZFC process has been magnified 200 times. (b) The evolution of superconducting transition temperatures ($T_c^{mag}$, determined by the separation point between ZFC and FC curves) with molar ratio $n$. Inset: Temperature dependences of zero-field-cooled (ZFC, closed symbol) and field-cooled (FC, open symbol) magnetization under 20 Oe along the $c$-axis for samples annealed at 400 °C and 600 °C with the same molar ratio $n = 0.1$. (c) Magnetic hysteresis loops (MHLs) for the AG and AVA ($n = 0.2$) samples at 5 K ($H \parallel c$). (d) Temperature dependences of normalized resistivity below 30 K for the AG and AVA ($n = 0.2$) samples.

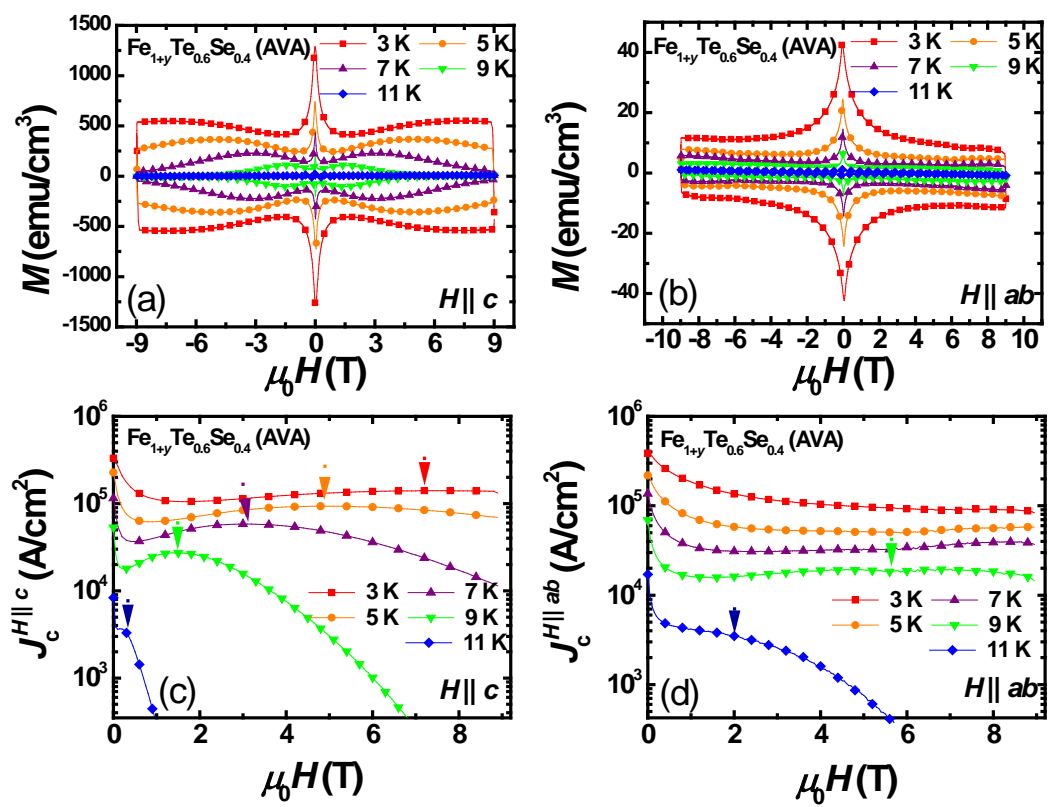

Fig. 2 (Color online) Magnetic hysteresis loops (MHLs) and critical current density ($J_c$) for one piece of AVA sample ($n = 0.2$) up to 9 T for fields applied both parallel to $c$-axis ($J_c^{H \parallel c}$) and $ab$-plane ($J_c^{H \parallel ab}$). $J_c$ values are calculated from MHLs using the Bean model.

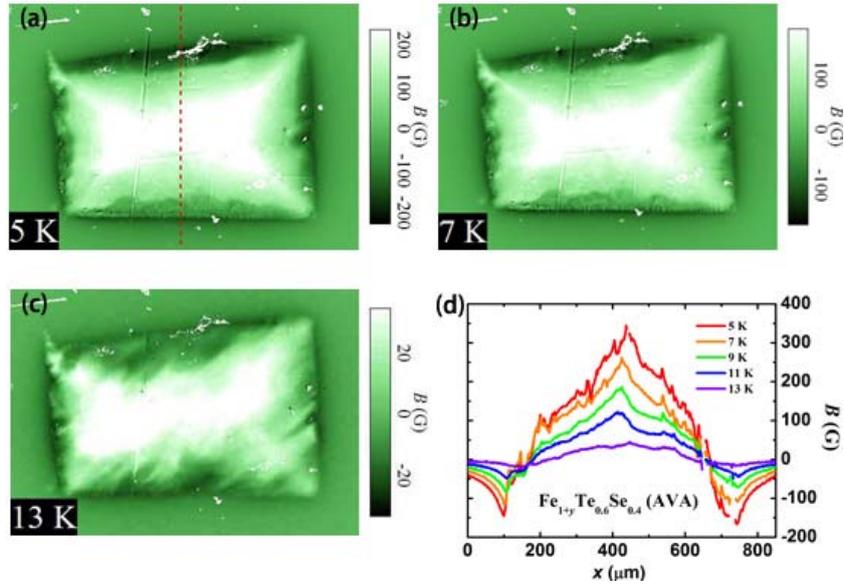

Fig. 3 (Color online) Magneto-optical (MO) images in the remanent state for a AVA ($n = 0.24$) sample at (a) 5, (b) 7, and (c) 13 K, respectively. (d) Magnetic induction profiles along the dashed line in (a) at different temperatures.

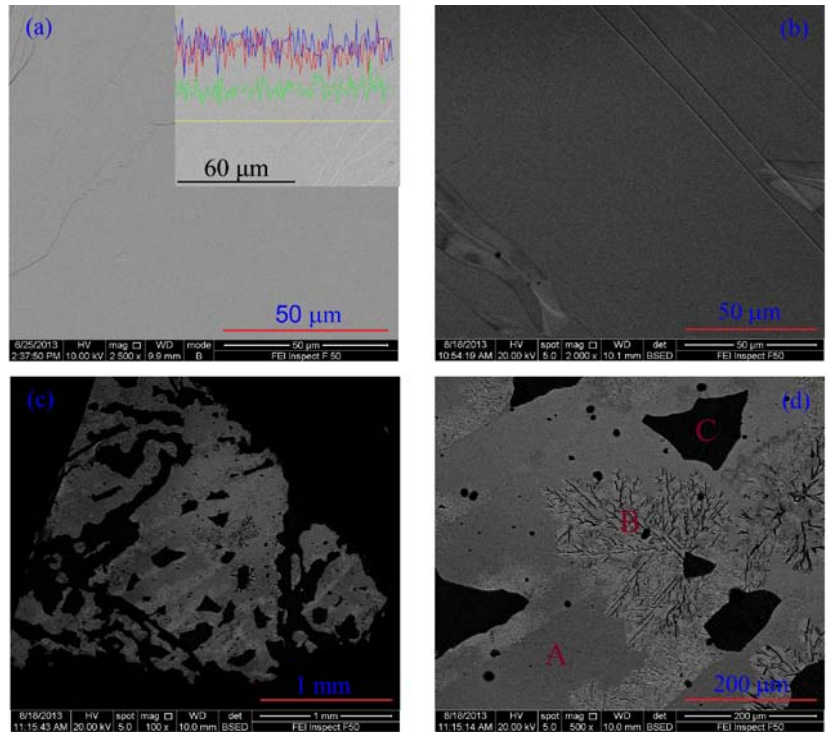

Fig. 4 (Color online) Back-scattered electron images of SEM measurement for the AG and AVA ($n = 0.2$) samples. (a) The topography of the AG sample. The inset at the upper-right corner is extracted from energy dispersive X-ray spectroscopy measurement. The blue, green, and red lines stand for the concentration of Te, Se, and Fe, respectively. (b) The topography of the inner part of the AVA sample. (c) The topography of the AVA sample without peeling the surface intentionally. (d) A enlarged view of

(c). Three different areas observed were marked with A, B, and C.

Table 1 Elemental analysis of the AG and AVA ($n = 0.2$) samples. The measured areas are marked in Fig. 4. The concentration of Fe has been revised by ICP results using the method stated in the main text.

| Area | Fe | Te | Se | As | Composition |
|---|---|---|---|---|---|
| (a) | 57.7 | 30.6 | 20.0 | / | $Fe_{1.14}Te_{0.61}Se_{0.39}$ |
| (b) | 54.4 | 31.9 | 20.0 | 0 | $Fe_{1.06}Te_{0.615}Se_{0.385}$ |
| (d)-A | 54.1 | 31.0 | 21.6 | 0 | $Fe_{1.03}Te_{0.59}Se_{0.41}$ |
| (d)-B | 42.8 | 35.5 | 23.2 | 3.9 | $Fe_{0.73}Te_{0.60}Se_{0.40}As_{0.07}$ |
| (d)-C | 35.7 | 2.8 | 0 | 65.9 | $FeTe_{0.08}As_{1.85}$ |